# ESA's Voyage 2050 Long-term Plan for Education and Public Engagement

## WHITE PAPER

**Corresponding author:** Pedro Russo
Astronomy & Society Group, Leiden Observatory, Leiden University, the Netherlands
P: +31654372658 E: russo@strw.leidenuniv.nl

# Table Of Contents

# Introduction

This white paper responds to the Voyage 2050 Call for White Papers[1] from the Science Programme of the European Space Agency (ESA[2]) and argues that education, communication and public engagement (hereafter EPE) should have priority in the Voyage 2050 planning cycle.

The ESA Science's Voyage 2050 missions promise insights into the big existential questions of our era: the prevalence of life in the Universe; the nature of space and time; and the intertwined nature of matter, energy and gravity. It is likely that innovations in the acquisition, handling and processing of vast data sets will drive these themes to scientific maturity in the next decades. They offer us a timely opportunity to underline the relevance of space sciences to everyday life and thinking.

More generally, space science is maturing to the point where it contributes to every major aspect of our cultural discourse. Citizens need information, resources and opportunities to actively participate in that discourse, and ESA Science can provide these.

This white paper is a modest attempt to support ESA Science improve its engagement with society. It focuses on issues and topics to improve ESA Science's Education and Public Engagement activities. It does not dwell on the topics that ESA already excels at; hence this White Paper provides a critical review of what should and could be improved. We believe ESA's Voyage 2050 programme teams have a responsibility to represent Europe's social and cultural diversity, and our suggestions are conceived in that spirit: to support ESA Science's complex task of engaging a hugely diverse audience in the complex issues of planning, building and operating fascinating space missions.

# Executive Summary

**In this White Paper we discuss aspects of education and public engagement relevant to ESA Science's Voyage 2050 programme. At the core of the paper is a set of recommendations from our expert group to support the planning, development and operating aspects of Voyage 2050 missions:**

We propose that ESA Science moves from Education and Public Outreach (EPO) to Education and Public Engagement (EPE) in both emphasis and terminology.

We recommend that ESA Science becomes an enabler of the EPE community in the same way that it is an enabler of the space research community, and we recommend the deeper involvement of the European EPE community at the mission level.

ESA Science should invest in a coordinated strategy for deploying embedded EPE teams to work alongside each Voyage 2050 mission. ESA's EPE strategy should make use of and support communication and collaboration between mission and national teams.

EPE should be fully recognised as part of the process of a Voyage 2050 mission and at least 2% of the budget of a mission's life cycle costs should be spent on EPE.

Rather than waiting to highlight a mission's scientific achievements, ESA Science missions should engage the public at the earliest stage of an ESA mission's life cycle, exploring the challenges a mission faces and the themes it will address.

EPE efforts should adopt a person-centered, narrative approach to ESA missions, in which the personal stories of the people involved are central to making the science enterprise relatable to the public.

ESA's EPE materials and resources should be properly translated and localised for each ESA member state.

ESA should participate in and contribute as much as possible to Europe-wide efforts in EPE, including European Union-funded projects.

**Public Engagement**

ESA Science should train the Voyage 2050 research teams in EPE skills. This training should involve representatives of minority groups to highlight the diversity of experts who are making the Voyage 2050 missions possible, and who provide a wide range of role models for future generations of space scientists.

ESA and its national offices should be resourced to collaborate with innovative forms of storytelling and work closely with diverse European Creative industries (e.g. music and cultural festivals, films, games, travelling exhibitions and even fashion).

ESA's programmes of education and public engagement should be based on its own research. Missions should be a proving ground for small teething and experimental projects in order to identify promising avenues for development.

ESA should develop a citizen-science strategy, collaborate with current programmes (e.g. Zooniverse[3], CSLab[4]) and co-develop new ones. Voyage 2050 missions should include citizen-science components in their programmes of data analysis.

ESA should further invest in fully open high-quality audiovisual materials (science images, artists' impressions, animations, 3D models) to illustrate missions, especially at their early, candidate stage.

**Formal and Informal Education**

ESA should devote further efforts to lobbying at a ministerial level, to embed space science content into existing national curricula in a cross-disciplinary fashion. We recommend that ESA's ESERO network (and other national contacts) lobby national education ministries for increased space science and astronomy content across the school curriculum, to link with its science missions.

ESA's direct educational efforts should focus less on adding to the classroom burden of teachers, and more on the informal learning opportunities afforded by individual missions.

ESA educational activities should follow international quality standards, both in terms of their accessibility as open educational resources, and in their being peer-reviewed through the IAU's astroEDU platform[5].

ESA should supplement traditional approaches to formal education (e.g. teacher training) with innovative efforts to engage with and provide lifelong learning opportunities for a large, diverse audience, regardless of its level of space-science literacy.

We recommend ESA specifically includes families as a key audience, and recognises that they are a major influence on young people's development.

Collaborating with major universities and informal learning institutions, ESA should use ESERO to establish venues for informal learning around Voyage 2050 missions. Resources should be shared with venues in other territories, who will then adapt them for their own audiences.

# ESA's Science Missions and Society

**ESA drives investment and growth across Europe and beyond as it addresses, from its unique perspective, topics that impinge directly on people's lives, from health and longevity to air pollution, from food security to climate change.**

Much of this work, relevant to people's day-to-day existence, does not enter public consciousness in a way that highlights ESA's role in its achievement. When knowledge acquisition and technology transfer are perceived as "natural goods" in this way, they become vulnerable to accidents of misattribution, acts of misinformation, and cultures of distrust. For the sake of a solid democracy, a better, more fertile, more durable connection should be made between ESA and its public. A more visible ESA and its achievements will, in any event, lead to greater public appreciation and support of its missions.

It is unfortunate, though inevitable, that ESA should struggle for public recognition in a post-space race world. Seen from the perspective of the 50th anniversary of the first human steps on the Moon this year, it is clear that the space race acquired a unique, unrepeatable cultural significance in an extraordinary time. For a while, NASA became a cultural icon.

But times change and icons fade. What should a space agency fit for this century look like, and how should it address and engage with the diverse communities that support it?

ESA's hybrid nature, encompassing elements of government office, research agency, and a popular venue for ideation and speculation, sets it, we believe, on an exciting, public-facing path.

### ESA's public communications

ESA public communications are a portal to the complex, collegiate workings of the international space community.

ESA is responsible for building and flying missions; the associated instrumentation and science are funded at a national and/or institutional level. ESA is by nature and constitution a collaborator and producer, bringing funds and talent together from many organisations distributed over many locations.

The very richness and scale of this culture is hard to contain in a story. The time-frame for reporting on a mission from conception to completion can span a generation. The scientific and technological contributors for each instrument carried by a mission may be scattered over the whole continent and beyond. For ESA to tell stories about who is doing what for a dozen or so missions in flight, up to twenty future and candidate missions, not to mention 40 years of past missions, is a mammoth task. Temporal, geographical, linguistic and cultural challenges abound.

ESA necessarily promotes itself as a single governing body to its sources of funding, to governments and ministries, and to allied institutions. It does not need to engage with the public in the same voice, if alternative strategies can be more effective or efficient.

Central to our white paper is the idea that some public engagement and educational opportunities fostered by each mission are best handled at the level of the mission itself. A more decentralised EPE effort has already been shown to be successful for NASA's science missions and ESA/Hubble efforts.

A clear strategy and support structure can be provided centrally from ESA. This will facilitate and empower the delegation of some of ESA Science's EPE activities to the teams and groups working on missions around Europe. This localisation effort will better leverage, in the different countries involved, the knowledge, insights and culture that evolve around each mission. This will free ESA from the difficult obligation to speak to all audiences all at once.

ESA has been for the last decades using mainly one-way communication and education approaches to communicate with different audiences. The most recent Science & Technologies Studies policy reports[6] recommend a deeper engagement with citizens, employing dialogue, co-creation of knowledge and participation. Some ESA initiatives have been moving in that direction; however, a more systematic implementation across the organisation is needed.

ESA Science's education and public engagement should reflect the cosmopolitan and collegiate culture it has spent so many years fostering. ESA Science should think of itself as a producer, whose productions (missions) involve many parties and interests.

People come to know and understand ESA through its missions, rather as cinema-goers come to know a company like Sony Pictures through the feature films it makes. Thinking of itself as a producer — a powerful partner and enabler of talent — will help ESA Science understand and make the most of new forms of communication that engage students, communities, and the general public.

**Towards a Mission-led communications strategy**

ESA is, of necessity, a complex organisation. The managerial level of the organisation, while vital, is not interesting to outsiders. ESA's philosophy, passion, and reason for being is reflected in the assets it brings to life; consequently, its EPE should focus on these assets.

These assets are both technical and human. They are, for example, the Philae lander itself, the people who first drew the lander on the back of a napkin, the university departments that overcame the challenges in building it, the companies that created and tested the unique parts and so on, along chains of research and development, testing and improvement.

ESA should find and focus on the stories triggered by its work. People outside the space community are interested, first and foremost, in what ESA does and the challenges inherent in accomplishing its missions.

A mission is a container for many smaller, more personal stories. At their best ESA Science's EPE activities reflect the distributed nature of mission planning, production and assembly by identifying and celebrating all the many local institutions, groups and individuals involved in the production of a mission. Because ESA so often attempts to speak to all audiences with one voice, its tone can seem to be both pro forma and peremptory, rather as though it is pushing to the front as the mission gathers pace. This single standard voice while having some benefits, it has the major disadvantage of speaking to all and none, in disengaging the public because the discourse has been stripped of all local, personal and human stories. ESA communication becomes an impersonal, uniformed, ready-made product with which the public feels no connection.

Currently, the communications strategy within ESA's Science Directorate is focused around mission milestones. Starting with the announcement of the call for mission concepts, through the phases of mission selection and mission construction, ESA Science provides periodic updates linked to milestones (the integration of key hardware, for example). Months, or even years, can pass between these events, and it is left to the (largely unsupervised) initiatives of mission and instrument teams to fill in the day-to-day story of how the mission is coming together.

We recommend increased investment and a more coordinated strategy for deploying embedded communications teams. Working alongside each mission, these teams are ideally placed to reflect a mission's constantly changing priorities and levels of activity. We recommend increased investment and a more coordinated strategy for deploying embedded communications teams. Working alongside each mission, these teams are ideally placed to reflect a mission's constantly changing priorities and levels of activity, and to develop best practice. The take-up of training within EU-funded programmes (e.g. Europlanet 2020 RI, ESConet) demonstrates the need within the research community to develop science communication skills.

It is currently left up to mission teams to commission visual assets to illustrate the earliest stages missions. Consequently, these assets vary in quality and style; frequently they don't exist at all. The lack of high quality visual resources robs ESA Science of opportunities to highlight early-stage missions in the context of the wider, global conversation about space (scientific discoveries that relate to mission targets, high-profile mission events by other space agencies, and so on). Putting at least some of these assets in place at the candidature stage would reap benefits over several years.

Currently, it is only around a year to six months before the launch that a full strategy for mission communication is developed and implemented within ESA Science. At this point, there may be close collaboration between all the actors involved. However, facilitating and supporting communication between ESA Science and the mission teams from the earliest stages of mission concept would lead to better practice and more effective messaging by mission teams. This in turn would foster a much broader, in-depth engagement between ESA and citizens across Europe.

Individual and specialist skills may be required for EPE work on different missions at different times. A certain amount of peripatetic working will be required, and this will cost money. A delicate balance must be struck between the amount of independence and decision-making at the central and distributed level. We would strongly urge against full centralisation of these efforts, however, as we believe the "embedded" nature of these teams is essential to their proper working.

Just as it operates R&D teams for its science and technology, ESA Science should establish an R&D team for innovation in communication, experimenting and adapting to new forms of media and content to support EPE mission teams. Analogous programmes include Fabrica's work for Benetton, MIT Medialab in MIT, and The Studio at JPL. These innovation labs should be formed and run in close collaboration with partners across Europe. An R&D team for educational innovation can also be established along the lines of NASA's Classroom of the Future[7], which produced highly innovative astronomy and space science educational products (e.g. Astronomy Village: Investigating the Universe, Astronomy Village: Investigating the Solar Systems, Exploring the Environment, BioBLAST). These products enjoyed wide dissemination and woke a generation of students to the value of earth and space science missions.

## Production of ESA communications

Much work has already been done to dismantle the barriers to effective collaborative communication by ESA Science and the mission teams.

The attempt to introduce effective Creative Commons licensing rules in 2014 addressed the way a team's ownership of data could delay and disrupt useful public communication and prevent tax-payers from using and benefiting from science results. However this important change has not permeated fully and and requires a buy-in not just from ESA itself but from the whole community that contributes to its missions.

EU funding via European Union's FP7 and Horizon 2020 has enabled ESA's past, current and future data archives to be more widely exploited. The EXTraS programme[8] was an early and exemplary example of what is possible[9].

ESA's most serious communications shortfall lies in the area of translation. ESA fails to translate its materials in a timely manner (or, indeed, at all) into the languages of the countries that fund it. This represents a major and persistent shortcoming. Thankfully, it is one we know how to address (witness the growth in translation capacity at the European Space Education Resource Office (ESERO) and the ESO's Outreach Network.). The infrastructure for improvement exists; Member states' communication teams in ESA national offices have all the necessary skills and what they lack is financial support.

Translations of real-time news into the major ESA languages (English, French, German) is efficient. "Smaller" language groups are less well served, and this is contributing to widening the gap of access to EPE resources between ESA member states. Even less excusable is the absence of translated content on the ESA's own website (e.g.: only 10% is available in Polish).

In the very worst cases, there is little to suggest that ESA is interested in collaborating with its own national offices. Translation services may be slow, scant, and even absent. Print and publication of ESA materials is uncoordinated. ESA often provides only a few generic English material and expects national offices to allocate resources to produce their own local copies. If ESA wants to be promoted in its member states it has to provide the necessary support, not just the demand.

Translation services cannot be simply conjured into being through a service contract. Reliance on voluntary work or funding from national office budgets is unacceptable. Ideally there should be a team of professional translators, funded by ESA who are given timely access to news, so that when it is made public, it is immediately available in all languages. National institutions who would have an intellectual stake in such work should be approached in each funding country.

Making information available to stakeholders is important. Making the right information available to the right audiences, and in a form they can access, is just as vital. Not everybody enjoys or has the time to read long, exhaustive press releases. These serve an important function, providing detailed information in a relatively handy form. They ought however to be the source from which simpler and more direct (not to say more engaging) communications are drawn. At the very least, they need a more usable structure, for instance by placing more fine-grained information in a "learn more" section.

To properly address this shortcoming, however, systemic change is required. Clear channels need to be established so that mission teams know how best to notify ESA of news, events and activities. The incoming mass of stories needs to be evaluated by nimble and active teams who propagate stories at different lengths and levels of detail through all ESA's media channels. Social media works at or close to real time, and ESA's current public offering feels very wordy and ponderous when transferred to this medium.

Today, no one can be said to "own" a public conversation. Channels of communication have proliferated in which every user can cast themselves as author, editor and/or publisher. In this environment, it is very easy for an organisation to end up contorting itself to fit its idea of what other people expect of it. This is expensive, exhausting, and never works. Timeliness and directness are much more realistic, effective and affordable goals.

More emphasis, and appropriate funding, should be placed on nimble photo and video production. Early stage missions lack the visual resources they require to communicate their work to the press and public. ESA's image archive also needs further improvement: it is still slow and unintuitive to operate and advance search does not work. Communications over social media need to be plentiful, timely and consistent. ESA should also make sure that communication plans developed at mission level have an issues management and crisis communication component. On a real-time medium like Twitter, silence, from a normally engaging source, is as eloquent as speech. Philae's problematic landing, live on Twitter, was a fascinating story that engaged a committed and positively minded audience. The Twitter silence that accompanied Schiaparelli's crash-landing on Mars was, by contrast, deafening.

# Public Engagement

**No scientific discipline, however specialised, exists outside the culture that sustains it. Even the most cursory review of its history reveals that astronomy speaks to philosophy and to music quite as much as it speaks to physics and engineering. Regardless of their space-science literacy, citizens should be given the opportunity to develop some broad cultural understanding and appreciation of ESA's work and purpose.**

Engagement at the popular level is not about conveying facts. It is about inspiring, shaping and deepening personal interests and explorations, in the expectation that people's grasp of the facts will follow.

Imagine how a discussion about alien ecosystems might inform public conversation about ecology generally; or how a description of radio pollution, and the way it is interfering with studies into black holes, might broaden how the public thinks about pollution more generally. In neither case does ESA have to own such a conversation. What it can and should do is bring its own experiences to the conversations it enters, and we believe ESA can do this most effectively through popular entertainment and media.

Addressing the scale and wonder of ESA's theatre of operations will generate more interest and curiosity than any amount of technical education. By sharing its mission insights, ESA has revealed that other worlds are possible. Some observers may draw close-to-home ecological lessons from what they have gleaned about ESA's work. Others may think about the place of life in the wider cosmos. Still others might think about the technical and human challenges we face in space.

### Engaging with popular culture

Drawing uninitiated audiences into one's own venues, to engage with one's own activities, is expensive and almost always fails. Far better, ESA Science should go to where the audience already is, collaborating with venues and institutions outside its field.

Embracing and engaging with popular culture is about establishing a simple shared vocabulary. It should be remembered that few European citizens have much substantive knowledge of space science. A rock band like Muse singing about supermassive black holes creates a better foundation for public engagement than any number of transdisciplinary fine-art shows addressing the well-educated and already-converted.

A pop concert may present concepts and moods in an engaging way. Participation, a more complex ambition, may be achieved through co-creation, sometimes involving the plastic and visual arts. Design fiction, guided speculation and other forms of co-creation practice place scientific and engineering content in a richer context, providing audiences with the tools necessary to participate in discussions. This could be built upon the the existing ESA partnership programme.

Engaging with well-known figures from the worlds of games, cinema, music and entertainment would present few challenges for an organisation as big, influential and

**Mission Narratives: A potential storytelling template for ESA's Voyage 2050**

*Phase 1: Begin with ideation and end with a launch*

From a narrative point of view the launch is the climax or midpoint of a mission, not its beginning, so people need to be involved in a mission long before its launch date.

Indeed, public engagement should start as soon after a mission's conception as possible. Missions do not have to be selected and approved before they are interesting to the public. Competition between proposals is the stuff of which stories are made, and unsuccessful proposals are, at the very least, prompts to the imagination. The ideas come first, and are there to be explored for ten, twenty, maybe thirty years before the results come in — assuming the mission is a success in the first place. We feel strongly that ESA Science needs to front-load its education programme so as much work as possible is done before a mission launches, in order to capitalise on the excitement and spirit of the new investigation.

*Phase 2: When the story falls silent (e.g. transit)*

Not all educational projects can be maintained in the long term, no matter how effective or innovative that they are. There is an important role for EPE professionals to think about how to plan sensible decay curves for projects as the support resources become more limited. In some cases, , the injection of minimal resources at key points in a project might extend or flatten the curve that describes the life and value of the project. Why is this a concern? By extending the lifetime of a project, much like spacecraft like the Voyagers have had their working lifetimes extended, we can maximise the total educational productivity with a minimal amount of additional resources.

*Phase 3: Findings, insights and assessments*

The chance to tell stories about transnational collaborations, inclusiveness and equity (or lack of same!) should not be disregarded. Positive examples of inclusiveness, diversity and equity should be highlighted in order to provide role models for future generations of truly diverse space explorers.

*Phase 4: Exploring the future*

ESA Science, in cooperation with wider communities, should openly present and discuss the implication of Phase 3 to the future and its next missions.

(frankly) as interesting as ESA. Major artists provide a bridge between the specialisms of space science and the concerns of the public. Environmentalism, climate, the possibility of life on other worlds, and concerns about pollution are likely to figure strongly.

A willing audience learns quickly. A confused or indifferent audience does not learn. ESA Science should conduct, in partnership with other European organisations (e.g. the science education and communication departments of European Universities) research into its audience, its production methods, and the kinds of stories it tells. It will need to learn the language of pop culture, and speak to audiences in terms that inspire their interest and curiosity.

## Handling failure

ESA Science should celebrate its culture of quality: how careful planning and testing can enable innovative investigations that are perceived by the public as risky and experimental. The public is excited by stories of how ESA's designers, engineers and astronauts manage and live with risk. ESA does not need to apologise for high-stakes missions and should stress the lessons learnt from failures or near failures. ESA stories in which there was some threat, challenge or problem to overcome should be of particular interest to the public.

Failure is a delicate topic. Communications around potential or real failure should be the result of an issues and crisis communication strategy set out far in advance. As part of a mission's EPE statement, ESA Science should require that such a strategy be proposed, and be further developed with ESA's central communication office when the mission is approved.

## Creating a culture of creativity

Every mission has a unique character. Every mission has a story to tell. Mission teams need at the earliest possible stage to discuss what kind of story a mission will tell, what its character is, and what its tone will be. The technology and engineering side of a mission may have just as many interesting challenges and stories as the science side.

The ideal ESA mission narrative holds in balance three main components:

- a human story
- a science question/investigation
- a technical puzzle

Depending on the mission, each of these aspects will at some point come to dominate the narrative.

Mission-based EPE should be delegated to embedded teams who understand their missions from the ground up. To a degree, this is already happening, though in a rather unplanned and sometimes disconcerting manner. The live-action short movie *Ambition*[10] with its dedicated campaign, made to promote and celebrate the 2016

Rosetta mission, is an eloquent case in point. An agile team sidestepped the centralised communications channels ESA had provided in order to invest a large budget in a piece of public communication which was then under-served by ESA's unprepared distribution infrastructure. *Ambition* was, artistically and in terms of communication, a stand-out success. But this form of working cannot be easily supported by an unprepared distribution channel, nor can it easily build institutional trust.

Creating a culture of creativity and risk taking is a challenge for science, engineering, and education teams. In some cases, short-term hires of creative designers and innovators is necessary to help design teams and managers to become more comfortable with innovation. In the case of the NASA Classroom of the Future, hiring highly creative programme managers and consultants helped jump start the development process.

### Engaging citizens in creating mission science

JunoCam, on board NASA's Juno mission to explore Jupiter's magnetosphere, is an excellent example[11] of how citizens can bring external expertise to contribute to and enhance the impact of the science derived from missions.

ESA Science has had some success in harnessing the very active amateur communities around Venus Express, Mars Express, Rosetta and, most recently, Gaia Alerts[12]. However, these have largely been grass-roots initiatives driven by the community. The role of amateurs and experts from beyond ESA's usual stakeholder groups has the potential to raise awareness and increase the likelihood of technology transfer both into and out of the project, so maximising its socioeconomic benefits.

Still a decade away from its launch in 2028, the ARIEL mission has run an international machine-learning competition[13] to improve the processing of data it expects to glean from its exoplanet observations. The challenge has attracted over 100 competing teams, making it one of the largest machine-learning competitions worldwide in 2019. This kind of activity creates recognition for ESA within new communities, as well as increased visibility in the wider media.

### Diversity and inclusion within public engagement

Public engagement with ESA can only come from public trust. Excellence alone cannot build this trust. ESA publicly recognises its responsibility to use its technology for the further development of humankind, and it shows through its actions that its programmes of research and innovation are not taking place in a social vacuum.

In 2015, the United Nations established Sustainable Development Goals, with the aim of ending poverty and hunger by 2030. These serve as vivid reminders of the connection between people and planet, and ESA has already developed a wide range of programmes to support these goals[14].

All public engagement and educational aspects of ESA Science missions should be designed in light of this work, to reflect the contributions each mission makes to development, diversity and inclusion. EPE is more than communication — it is the route by which communities and individuals outside the specialisms of space science, and who fund and stand to benefit from that science, help define, implement and assess the missions ESA Science produces.

The alliance to be made between large-scale citizen input and expert knowledge is a powerful one, as evidenced by the outcomes of EU-funded agenda-setting projects such as Voices[15] and Cimulact. Novel techniques and approaches are required to facilitate a dialogue between specialists and citizens, but plentiful examples of good practice exist to guide this effort, for example the EU Seventh Framework for R&I (FP7)[16]

These projects also demonstrate the importance and efficacy of concerted engagement with under-represented communities (whether defined by age, class, socioeconomic background or other characteristics), including specific actions to challenge and counter discrimination[17]. ESA Science's efforts here must be institutional, long-term, and robust enough to navigate fads, trends and capture by vested interests.

First, ESA's materials should be properly translated and captioned and available in multiple formats. This is a production problem, not an editorial problem, and requires additional resources. The investment on these resources will pay back in terms of more awareness about ESA among the public and, more importantly, by reaching talented individuals who can make key contributions to space science, but who would otherwise be excluded because of their disability.

Next, public awareness of space science will be immeasurably improved if information is provided through multiple channels. People retain information much more readily if they glean it from multiple sources, and over varied media. In this way, using multiple media promotes inclusion. Disabilities shape a person's uptake of different media. Multiple channels not only ensure effective communication; they engage otherwise marginalised audiences.

Virtual Reality(VR) and Augmented Reality(AR) experiences provide an excellent demonstration of this point. Projection-mapping domes — themselves inspired by planetariums — can be used to stage live events from space. Footage like that shot in the cupola of the ISS in 2014[18] can bring immersive experiences of space to the public using VR, AR and projection-mapping technology. VR and AR experiences are also well suited to be accessible to persons with special needs (someone in a wheelchair is able to move through a spaceship; a blind person can listen to the sounds and the descriptions of what she is encountering; a deaf person can read brief information).

# Formal and Informal Education

**ESA Science is well placed to be an innovation leader in education. It is trans-disciplinary to its core. It engages with citizens both within and beyond formal education. The timeframes of its missions define what is meant by lifelong education.**

The information revolution has thrown up new, more inclusive strategies for learning. The old "deficit" model of education is ceding ground to more creative, playful forms of address. Now that YouTube is the biggest public education platform on the planet, ESA must must abandon the sureties of the past and use new tools to create its own, experimental future.

**Formal Learning**

Education is local. It is culturally specific. As the project Space Awareness (EU-SPACE AWE)[19] has demonstrated, research into local curricula is vital, and can best be done at a local level, through ESA's national offices. ESERO national offices research national curricula closely, acquiring the level of fine-grained local detail they need to lobby effectively for space education at a ministerial level. Prior to approval each ESERO candidate conducts a study to see how best to include space topics in the national curriculum.

ESA Science focuses its educational efforts on teacher training, and has invested heavily in this area. ESERO is ESA's main way of supporting the primary and secondary education community in Europe. We note, however, how hard it is to maintain conversations with the educators ESERO instructs and inspires.

Consider the relatively poor uptake there has been for the European Astro Pi Challenge[20]. This project offered young people the chance to conduct scientific investigations in space by writing computer programmes that run on Raspberry Pi computers aboard the International Space Station. The opportunity was real but few students and schools rose to the challenge, as it sat outside the national curricula those students and schools had to follow. While it would be easy to criticise ESA Science for teaching only what it was interested in, the actual picture is more complicated. The absence of coding in some national curricula hampers ESA Science's educational efforts, and cannot be addressed simply through teacher training.

Teachers need support covering their curriculum day to day. Working alongside teachers in support of the curriculum will maintain lines of communication and goodwill with the education community. We suspect that ESERO achieves more when it works directly with students on curriculum enrichment activities, than when it provides teacher training programmes which teachers do not have time to implement in the classroom.

ESA Science should avoid mundane or "stamp collecting"-type science education activities, and focus on areas that throw up innovations and serve new audiences.

The outcome of such efforts is never predictable, and the culture of ESA Science's EPE organisations at research centres ought therefore to be congruent with ESA's overall culture of R&D.

As in scientific research, some projects or activities should be highly experimental and carry a risk of failure if they are to provide practitioners and participants with a significant opportunity for learning and growth. Indeed the values of risk and failure should be discussed within the activities themselves, to counter the prevailing impression so often given to students that the world has already been explained and explored.

All ESA Science educational activities should follow international quality standards, both in terms of their utility as open educational resources (OER), and by being peer-reviewed through the IAU's astroEDU platform[21]. OER are freely accessible, openly licensed text, media, and other digital assets that are useful for teaching, learning, and assessing. OER provide an open license so any user can use, re-mix, improve and redistribute the materials under some conditions. IAU astroEDU is an open-access platform for peer-reviewed science education activities, where all the activities are peer-reviewed by two referees; a research scientist and an educator. This process provides activities with heightened credibility by increasing the quality of the scientific content and educational implementation.

**Informal learning**

Learning is also taking place outside the classroom, schools are being connected with local communities, with scientists, science centres and museums, industry and civil society, therefore creating new opportunities for students to learn at an individual pace and according to each child´s own interest and abilities. Informal learning also enables citizens to continue learning throughout their lives, acquiring useful work and life skills that will ensure they can continue to play an active role in society.

We believe that ESA should tap into the enormous potential of non formal education and lifelong learning by engaging with science centres and museums, citizen science organizations and organizing citizen´s debates in member states more often.

ESA Science should engage directly with citizens of all ages across all ESA member states through citizen-science projects and the wider informal learning sector, reflecting not only its work on missions and their scientific achievements, but equally its engagement with a wider culture that includes explorers, researchers, entrepreneurs, creators, storytellers and other societal role models.

Since 2012, ESA has developed an effective mechanism for collaboration with visitor attractions by founding a space working group within the Ecsite network. Ecsite's membership includes over 350 organisations in Europe and worldwide (including science centres, museums, planetaria, research bodies, festivals and companies) that have a collective reach of around 40 million citizens each year. Through the Ecsite

Working Group[22], ESA has created exhibition materials and informal learning resources relating to its missions and activities, including Rosetta, the ISS and lunar exploration. This working group provides a template for co-creation and collaboration between ESA and communications professionals as discussed elsewhere in this document.

Citizen Science Alliance/Zooniverse (Oxford University) is the world's largest platform for people-powered research. Volunteers come together to assist professional researchers. This type of citizen science is highly relevant to space science — indeed it grew out of the Galaxy Zoo project. However, to date there has been relatively little application of ESA Science missions or data within Zooniverse or other similar platforms. We recommend that all missions investigate the opportunities for citizen science projects from the earliest stages of planning. Hubble Asteroid Hunter and Euclid — Challenge the Machines[23] show that there are opportunities to involve and engage the public in missions through this type of informal learning, even several years before launch.

**Mission-based learning**

Space activities in general, and the distributed, collegiate, collaborative mission designs of ESA Science in particular, have great potential to inspire young people, foster critical thinking, and encourage a feeling of global citizenship.

ESA Science's missions may be used to develop part of the national curricula in many disciplines. It may even be possible to subdivide a mission into very small educational units — 1-3 activities each — so to cover the whole spectrum of national education, from science to maths, physics to art, philosophy to chemistry, history to music, depending on national requirements and the target group.

The education programme should feature many aspects of a mission or program, and not just the science. The term STEM is, after all, an invitation to integrate subject areas. Achievements in each area contribute to some overall goal. The challenge of STEM education is to show how these areas connect.

An EPE programme may also incorporate core corporate culture issues, for example the issue of quality control and the space industry's culture of risk analysis and mitigation. A project that has manufacturing challenges might feature modern manufacturing and the technical aspects (e.g. 3D printing, materials testing, inspection.)

Imagination and experiment sit at the heart of both the arts and sciences, and communicating ideas across cultures is best done in spaces where speculation and imagination are encouraged. We therefore recommend the use of a co-creation model in which experts, the public and youth committees shape the wording, spirit and social context of the questions ESA Science missions set out to answer, and the forms of engagement ESA develops around those missions.

This methodology is about more than words. Speculative design is used to give ideas form and purchase on the real world. Participants are invited to come up with ideas, think them through, and consider what the world would be like if their ideas were adopted. Consequently, participants acquire literacy in unfamiliar disciplines — from ethics to aesthetics to astrophysics — in a practical context, without encountering formal barriers to entry.

**Diversity, inclusion and equity in education**

ESA Science should tackle tangible barriers to inclusion in education through the better funding and management of schemes to translate, tour and license (or, better, open-source) its educational offering.

Breaking down these tangible barriers is important, and the first step in addressing problems of marginalisation and exclusion. ESA Science should also endeavour to address those intangible barriers that stand in the way of inclusion. We urge ESA Science to systemic efforts, rather than special ones. For example, ESA Science's mission teams employ a community that is admirably diverse on many axes; ESA's EPE, however, is not personalised so as to reflect (let alone enrich) that diversity. The encouragement of diverse role models always runs the risk of tokenism, and the proliferation of special siloed events, "weeks" and "seasons" can be a symptom of institutional anxiety rather than the result of managed change. Our proposal is conservative: delegate EPE efforts to embedded mission-led teams to encourage informal, diverse and (frankly) more interesting patterns of communication and dialogue.

# Conclusions

ESA is a highly distributed multinational organisation handling complicated communications and engagement challenges in an increasingly fractured and flattened media environment.

At the same time ESA Science, as educator, contends with the conservatism of national curricula, often straitened national education budgets, and a European cultural consensus that front-loads education on the young while providing little by way of opportunities for lifelong learning.

But innovations in media are already transforming our ideas about education, and by embracing these changes, and adopting the many innovations mentioned in this paper — from informal learning to co-creation, from real-time narratives to immediate and immersive media — ESA can turn the educational and media challenges it faces into exciting opportunities.

Some cultural changes are advisable. ESA Science can communicate more effectively with its communications and education stakeholders if it develops a range of suitable forms of address. Using the mission as an organising template, ESA Science can harness its teams, collaborators, friends and fans to make more visible its important role in investigating human existence in the cosmos. Despite spending considerable amounts of funding on EPE issues, ESA is nowhere near spending the recommended 2% of the budget on EPE in its Science programme.

The recommendations in this white paper are couched in terms of education and public engagement. However, we see this as the beginning of a conversation with ESA Science, leading to the deeper involvement of the European Education and Public Engagement community in the work ESA does to unlock the mysteries of the Universe.

# Endnotes and References

[1] www.cosmos.esa.int/web/voyage-2050

[2] We use "ESA Science" as short name for the Science Programme of the European Space Agency (ESA) and ESA as the overall Agency.

[3] www.zooniverse.org

[4] www.universiteitleiden.nl/citizensciencelab/citizen-science-lab

[5] https://astroedu.iau.org

[6] Royal Society, *Survey of factors affecting science communication by scientists and engineers*. https://my.su/87cz

[7] www.nasa.gov/audience/foreducators/postsecondary/internet/COTF.html

[8] www.extras-fp7.eu

[9] http://sci.esa.int/xmm-newton/60533-students-digging-into-data-archive-spot-mysterious-x-ray-source

[10] www.youtube.com/watch?v=H08tGjXNHO4

[11] www.missionjuno.swri.edu/junocam/

[12] http://gsaweb.ast.cam.ac.uk/alerts/about/

[13] https://ariel-datachallenge.azurewebsites.net/ML

[14] www.esa.int/Our_Activities/Preparing_for_the_Future/Space_for_Earth/ESA_and_the_Sustainable_Development_Goals

[15] www.voicesforinnovation.eu

[16] https://ec.europa.eu/info/research-and-innovation/strategy/support-policy-making/support-eu-research-and-innovation-policy-making/evaluation-impact-assessment-and-monitoring/past-framework-programmes_en

[17] https://ec.europa.eu/europeaid/sectors/human-rights-and-governance/democracy-and-human-rights/anti-discrimination-movements-0_en

[18] https://youtu.be/vuN_ItXMC-E

[19] www.space-awareness.org/en/

[20] https://astro-pi.org/

[21] https://astroedu.iau.org/en/

[22] www.ecsite.eu/activities-and-services/thematic-groups/space-group

[23] https://www.zooniverse.org/projects/hughdickinson/euclid-challenge-the-machines/about/research

# Authors' Biographies

**Pedro Russo** is professor in Astronomy & Society at Leiden University, the Netherlands. He coordinates the Astronomy & Society Group at Leiden Observatory, which implements global-scale projects in astronomy, space education and public engagement.

**Łukasz Alwast** is an honorary research associate at the Institute for Innovation and Public Purpose, University College London (UK), and chief development officer at Science Now, a design studio for science & technology communication, research and development.

**Lars Lindberg Christensen** is an award-winning astronomer and science communicator, responsible for the communication for some of the world's largest and most famous telescopes, including the Hubble Space Telescope (for ESA), and ESO's Extremely Large Telescope.

**Ewine van Dishoeck** is president of the IAU and a professor of molecular astrophysics at Leiden University, the Netherlands, studying star-forming clouds to planet-forming disks. She has received many awards, including the 2000 Dutch Spinoza award, the 2015 Albert Einstein World Award of Science, and the 2018 Kavli Prize for Astrophysics.

**Urban Eriksson** is an associate professor with specialization in astronomy education research at Kristianstad University, Sweden, and Chair for the global IAU Commission C Working Group for Astronomy Education Research and Methods.

**Edward Gomez** is an astronomer and education director for Las Cumbres Observatory. He is one of the founders of IAU astroEDU, a peer-review platform for astronomy education activities, and chair of IAU working group for astronomy education resources.

**Jorge Rivero Gonzalez** is global coordinator of IAU100, the centenary celebrations of the International Astronomical Union. He has also worked for the European Physical Society, organising the UN International Year of Light and Light-based Technologies 2015 and the UNESCO International Day of Light.

**Anita Heward** is communications manager for the ARIEL Space Mission and UCL's Centre for Space Exochemistry Data and is responsible for communications and outreach for the EU-funded Europlanet 2020 Research Infrastructure.

**Mairéad Hurley** holds a PhD in astronomy from Dublin City University and is education and learning manager at Science Gallery Dublin. She is principal investigator for SySTEM 2020 which studies informal science learning across Europe.

**Veronika Liebl** is director of organisation and finance at the Ars Electronica festival in Linz, Austria. Ars Electronica festival is a gathering of artists, scientists and technologists, intended as "a setting for experimentation, evaluation and reinvention".

**Ana Noronha** is the executive director of the Portuguese National Agency for Scientific Culture (Ciência Viva). Ana is co-chair of the ECSITE's Space Group and member of the ESA's Education Advisory Committee.

**Amelia Ortiz-Gil** is an astronomer and public outreach officer at the Astronomical Observatory of the University of Valencia, Spain. She was awarded the Europlanet Prize for Public Engagement 2019.

**Jan Pomierny** is chief executive officer at Science Now and founder of the design and production studio Stellar Fireworks. He produced the short feature *Ambition* (2014) for ESA's Rosetta mission and directed *Rosetta VR* experience (2016).

**Stephen Pompea** is a visiting professor at Leiden University, heads the US National Optical Astronomy Observatory's Education and Public Outreach program, and was named NOAO's first Observatory Scientist in 2014.

**Stefano Sandrelli** is an astrophysicist and head of outreach and education at the Italian National Institute of Astrophysics (INAF). He holds an annual course at the Master of Science Communication at the University of Milano, Bicocca.

**Oana Sandu** is a communication expert for science and technology who has coordinated award-winning outreach and communication campaigns for the European Southern Observatory, ESA/Hubble Space Telescope, the Romanian Space Agency, the Space Generation Advisory Council and others.

**Simon Ings** is arts editor of New Scientist magazine and writes about the intersection of art and science for the Financial Times, the Spectator and others. His most recent non-fiction title was Stalin and the Scientists (2016).

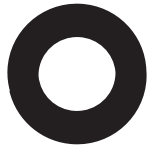